\null
\magnification=\magstep1
\hsize=16.2truecm
\vsize=23.5truecm
\voffset=0\baselineskip
\parindent=1truecm
\tolerance=10000
\baselineskip=17pt
\voffset=0\baselineskip
\parskip=0.5truecm
\par
 
\def\al{\alpha}
\def\be{\beta}
\def\ga{\gamma}
\def\c{\nabla}
\def\cc{\tilde\nabla}
\def\d{\partial}

\def\de{\delta}
\def\e{\epsilon}
\def\De{\Delta}

\def\gg{\tilde g}

\def\la{\lambda}
\def\La{\Lambda}

\def\oo{\over}
\def\o{\omega}

\def\si{\sigma}

\def\ta{\tau}
\def\~{\tilde}
\def\^{\hat}
\noindent
{\bf KERR-SCHILD METRICS REVISITED I.
THE GROUND STATE}\footnote{$^\dagger$}{Research
supported by OTKA fund no. 1826}\hfil \vskip.05in \noindent
{\sl L\'aszl\'o \'A. Gergely\footnote{$^\ddagger$}{Present address:
{\it Research Group on Laser Physics of the
Hungarian Academy of Sciences, H-6720 Szeged, D\'om t\'er 9, Hungary}}
and Zolt\'an Perj\'es}\hfil
\vskip.0in
\noindent
\line{Central Research Institute for Physics\hfil}
\line{H-1525 Budapest 114, P.O.Box 49, Hungary\hfil}
\vskip.05in
\noindent
{\bf ABSTRACT}
\midinsert
\baselineskip=12pt plus.1pt
   The Kerr-Schild pencil of metrics $g_{ab}+\La
l_al_b$ is investigated in the generic case when it maps an
arbitrary vacuum space-time with metric $g_{ab}$ to a vacuum space-time.
The theorem is proved
that this generic case, with the field $l$ shearing, does not
contain the shear-free subclass as a smooth limit.
It is shown that one of the K\'ota-Perj\'es metrics is a solution in
the shearing class.
 
\endinsert
\baselineskip=17pt
\voffset=0\baselineskip
\par
 
{\bf 1. INTRODUCTION}\par
 
  The challenge of Kerr-Schild metrics
for researchers in general relativity appears unabated for many
years now.  The particular way
Kerr-Schild metrics incorporate a congruence of null
curves in space-time geometry is a sure source of the fascination.
And then, an eminent member of this class is the Kerr black hole.
A brief critical review of the literature below will help us
through the ups and downs of the subject.
 
    The original Kerr-Schild Ansatz maps$^1$ Minkowski space-time
with a null field $l$, to an empty curved space-time with a metric
quadratic in $l$.
Boyer and Lindquist$^2$ show that the image of the map is an
algebraically special space-time in which
$l$ is a multiple principal null direction of the Weyl
tensor. Hence $l$ is tangent to a congruence of null geodesics.
The problem has been investigated in detail by Grses and
Grsey$^3$, who derive many of the properties of the resulting empty
space-time. Their treatment is unduly cumbersome though, as they
introduce a large number of various new fields.
\vfill\eject\noindent
\parskip=0.8truecm
 Another important contribution to the
subject by Debever$^4$ employs a Newman-Penrose (NP)
null tetrad adapted to the vector
field $l$. He normalises the
tangent vector $l$ of the null geodesics such that the parameter
is not affine. Unfortunately, however, his definition of
the shear, a key notion in the paper, is the one used
elsewhere in the affine gauge. McIntosh$^5$ and McIntosh and
Hickman$^6$ have considered the conditions that a Kerr-Schild
vacuum have twistfree principal null directions and to be of
Petrov type D. For a review of the original Kerr-Schild Ansatz,
cf. Chandrasekhar$^7$.
 
   Remarkably, even the general solution of the original
Kerr-Schild problem is known$^8$. The solution is provided by the Kerr
theorem, stating that there is a one-to one correspondence between
geodesic and shear-free null congruences in Minkowski space-time
and zeroes of analytic functions in three complex variables.
 
   The status of the original Kerr-Schild {\sl Ansatz} and its
generalisation where both the base space-time (to be named {\it
parent} space-time) and the total space-time are curved is
reviewed in the Exact
Solution Book$^9$. However, the subject appears to be plagued by false
statements. It is claimed in connection with the works of
Thompson$^{10}$ and Dozmorov$^{11}$,
that the vector $l$ must be a multiple null principal
vector of the curvature in both space-times$^{12}$.
(There is, moreover, a list of corrections issued by the authors
of the book which advises to cancel the paragraph on
double Kerr-Schild metrics in Sec. 28.5.)
Thompson$^{10}$ has obtained various results
on the generic pencil. Thus he has proved that the
null vector field $l$ of the pencil is geodesic provided
it is a geodesic of the parent space-time. For the vacuum-vacuum
Kerr-Schild maps, Xanthopoulos$^{13}$
has proven that the full field equations are implied by the
equations linear in the pencil parameter $\Lambda$.
Urbantke$^{14}$ has derived
Kerr-Schild solutions in alternative theories of gravitation.
 
   Recently, Nahmad-Achar$^{15}$ has investigated the types of
energy-momentum tensors which can be generated from an arbitrary
space-time by the
generalised Kerr-Schild map. By using the NP approach, he proved
that the ensuing space-time is algebraically special whenever the
parent space-time is an algebraically special vacuum.
 
   Despite these developments, there remain some tantalising
questions about the Kerr-Schild metrics. The original motivation for
our work was {\it question (Q1)}: Is it possible to construct
a `Fock space' for vacuum gravitational fields by creating a
multitude of Kerr-Schild congruences on some fixed vacuum base space-time?
We propose that the answer is no. This conclusion is based on our
research on {\it question (Q2)}: what is the condition on a
vacuum space-time that it can contain a
Kerr-Schild congruence mapping to a vacuum space-time. In this paper
we find a restriction on the
parent space-time of the vacuum-vacuum Kerr-Schild map. This
may be concisely put such that the generic Kerr-Schild map does
not arise as a smooth extension of the shear-free class.
Our result {\sl casts a serious doubt on
the chances that attempts relying on the Kerr theorem to carry
over to general relativity a complex-manifold description of Minkowski
space time can be successful.}
 
   In Sec. 2, we obtain the set of three field equations of the
vacuum-vacuum Kerr-Schild map. One of these is the condition that
$l$ is a geodesic vector field, a fact already known to Thompson$^{10}$.
In Sec. 3, we present the spin-coefficient form of the
pencil equations in the affinely parametrized gauge. Hence we
get the affine-parameter dependence of the field
quantities $\rho,\sigma$ and $\Psi_0$.
Sec. 4 contains our main theorem, the proof of which employs the
set of coupled equations for the spin coefficient fields
$\al,\be,\pi,\tau$ and $\Psi_1$. It follows from our theorem that
in the generic case, this set of equations is homogeneous and
linear. In Sec. 5. we get the `ground solution' of the general
Kerr-Schild map by taking the trivial solution of this set and
of the coupled homogeneous linear equations for $\la,\mu$ and
$\Psi_2$. The solution is then shown to be one of the
K\'ota-Perj'es metrics, by use of an Eddington-type coordinate
transformation.
 
\bigskip
{\bf 2. VACUUM CONDITIONS}
 
  Let $g_{ab}$ be the metric of a vacuum space-time and
$\gg_{ab}$ a pencil of vacuum metrics of the Kerr-Schild form
$$\gg_{ab}=g_{ab}+\La l_a l_b \quad\quad l_a l^a=0\eqno(2.1)$$
where the real constant $\La$ is the pencil parameter and $l^a$ is
a null congruence.
In consequence of (2.1) we have
$$\gg^{ab}=g^{ab}-\La l^a l^b
\qquad l^a=\gg^{ab}l_b=g^{ab}l_b \quad\quad \gg= g\ .\eqno(2.2)$$
 
  The covariant derivatives $\cc_a$ and $\c_a$ annihilating
$\gg_{ab}$ and $g_{ab}$ respectively, have the difference tensor
$C^c_{\ ab}$. The Ricci tensor $\tilde R_{ac}$ may be computed by
use of the relations$^{16}$
$$\eqalign{&\cc_a\o_b-\c_a\o_b=-C^c_{\ ab}\o_c \cr
&C^c_{\ ab}={1\oo 2}\gg^{cd}\left(\c_a\gg_{bd}+\c_b\gg_{ad}-
\c_d\gg_{ab}\right)\cr
&\tilde R_{ac}=R_{ac}-2\c_{[a}C^b_{\ b]c}+2C^e_{\ c[a}C^b_{\ b]e}}
                                                  \eqno(2.3)$$
where $R_{ac}$ is the Ricci tensor of the metric $g_{ab}$, defined
in terms of the Riemann tensor, $R_{ac}=R_{abc}^{\quad b}$.
The Ricci tensor $\tilde R_{ac}$ is a polynomial of degree 3 in
$\La$.
As the vacuum Einstein equations $\tilde R_{ac}=0$ hold for all
real values of the pencil parameter $\La$, we can equate
the coefficients of each power of $\La$ with zero. We get four tensor
relations, but one of them is satisfied identically. Thus we are
left with the equations:
$$\c_b\bigl[\c_a\left(l_c l^b\right)+\c_c\left(l_a l^b\right)-
\c^b\left(l_a l_c\right)\bigr]=0                  \eqno(2.4)$$
$$\textstyle
(\c_b l^b) l_{(a}Dl_{c)}+{1\oo2}(Dl_a)
(Dl_c)+l_{(a}DDl_{c)}+l_a l_c(\c_b l_d)(\c^{[b}l^{d]})-(Dl^b)(\c_b
l_{(a})l_{c)}=0                                   \eqno(2.5)$$
$$l_a l_c(Dl_b)( Dl^b)=0                          \eqno(2.6)$$
where the tensor notation refers to the parent space-time and
$D=l^a\c_a$.
From (2.6), the vector $l^a$ satisfies the geodesic condition:
$$Dl^a=f l^a                                      \eqno(2.7)$$
Using (2.7) and the Ricci identity
$$(\c_a\c_b-\c_b\c_a)l^c=R^c_{\ dab}l^d     \eqno(2.8)$$
we get from (2.5) and (2.4):
$$\textstyle
f\c_b l^b+{1\oo 2}f^2+Df+(\c_b l_d)\c^{[b}l^{d]}=0\eqno(2.9)$$
$$\c^b\c_b\left(l_a l_c\right)+2R_{abcd}l^b l^d-
2\c_{(a}\bigl[l_{c)}\left(\c_b l^b+f\right)\bigr]=0.\eqno(2.10)$$
 
  Equations (2.7), (2.9) and (2.10) ensure that a Kerr-Schild
pencil will generate a vacuum space-time from a vacuum space-time.
For future reference, we now rewrite these equations
in a Newman-Penrose form$^{17}$, choosing $l$ a vector of the null
tetrad $(l, n, m,\bar m)$. We choose the phase of the complex
vector $m$ such that the spin coefficient $\e$ is
real,
                                  $$\e=\bar\e\ .\eqno (2.11)$$
We can still perform spatial rotations
                          $$\tilde m=e^{i\Phi}m, \eqno(2.12)$$
with $D\Phi=0$, as well as arbitrary null rotations about $l$,
$$\eqalign{&\tilde l=l \cr &\tilde m=m+El \cr
&\tilde n=n+\bar E m+E \bar m+E\bar E l   \ . }  \eqno(2.13)$$
The geodesic condition (2.7) becomes
                     $$\kappa =0, \qquad f=2\e\ .\eqno(2.14)$$
Eq.(2.9) takes the form
$$\textstyle 2D\e={1\oo2}(\rho-\bar\rho)^2+2\e(\rho+\bar\rho)
-4\e^2         \ .                              \eqno(2.15)$$
The tetrad components of Eq.(2.10) are
$$\eqalignno{
\Psi_0&=-\si (4\e+ \rho-\bar\rho)                &(2.16.a) \cr
4\e(\rho+\bar\rho)&=(\rho+\bar\rho)^2-2(\rho\bar\rho+\si\bar\si)
                                                 &(2.16.b) \cr
D(4\e-\rho-\bar\rho)&=-8\e^2+2\e(\rho+\bar\rho)
-2(\rho\bar\rho+\si\bar\si)                      &(2.16.c) \cr
D\ta-\bar\de\si+\de\bar\rho&-4\de\e-2\Psi_1=-4\e(\ta-\bar\al-\be) \cr
&-\si\bar\ta+\si(\al+5\bar\be)
-\rho\ta+(2\rho-\bar\rho)(\bar\al+\be)+\pi\si+\bar\pi\rho
                                                 &(2.16.d) \cr
D(\ga+\bar\ga)-&2\De\e+\De(\rho+\bar\rho)-\de(\al+\bar\be)
-\bar\de(\bar\al+\be)-2 Re\Psi_2=\cr
&\pi(\ta+\bar\al+\be)+\bar\pi(\bar\ta+\al+\bar\be)-
4\e(\ga+\bar\ga)-2\e(\mu+\bar\mu)\cr
&-\si\la-\bar\si\bar\la-\rho\mu-\bar\rho\bar\mu
-\ta(\al+\bar\be)-\bar\ta(\bar\al+\be)\cr
&+4(\al+\bar\be)(\bar\al+\be)-(\al+\bar\be)(\bar\al-\be)
-(\al-\bar\be)(\bar\al+\be)                   \ .&(2.16.e)}$$
 
  These relations are complemented by the vacuum NP equations
(NP 4.2), the Bianchi identities (NP 4.5) and the commutators
(NP 4.4) of the four derivative operators $D,\De,\de$ and
$\bar\de$.
Equations (2.15) and (2.16.c) turn out to
be consequences of (2.16.a), (2.16.b), (NP~4.2.a) and (NP 4.2.b).
Hence, Eq. (2.5) follows from Eq. (2.4).
Moreover, Xanthopoulos$^{13}$ proves that also the geodesic condition
(2.6) follows from (2.4).
 
The task of solving the field equations is, however, better served
by a different gauge in which an affinely parametrized tangent vector
$l'$is selected for the Kerr-Schild congruence. This gauge will be
used in the subsequent sections.

\bigskip
{\bf 3. THE GAUGE WITH AFFINE PARAMETRIZATION}
 
  The Kerr-Schild map (2.1) may be postulated in the slightly
different form
$$\tilde g_{ab}=g_{ab}+V l'_a l'_b          \eqno(3.1)$$
where $V$ is a scalar function. Reparametrizing
by $$l=\phi l'\ ,                           \eqno(3.2)$$
where $\phi$ is a real function, the pencil (2.1)  becomes (3.1)
with
$$V=\La\phi^2       .                       \eqno(3.3)$$
 
  The scale of $l'$ is arbitrary in (3.1), and we fix it by
adopting an affine parametrization
$$D'l'_a=0  \ ,                                \eqno(3.4)$$
where $D'=l'^a\c_a$.
The rest (2.4) and (2.9) of the Kerr-Schild pencil equations
take then the form
$$\c_b\bigl[\c_a\left(Vl'_c l'^b\right)+\c_c\left(Vl'_a
l'^b\right)-\c^b\left(Vl'_a l'_c\right)\bigr]=0   \eqno(2.4')$$
$$D'D'V+(\c^a l'_a)D'V+2V(\c_b l'_d)\c^{[b}l'^{d]}=0.\eqno(2.9')$$
Some useful spin coefficients transform under (3.2) as follows:
$2\e=D'\phi,\quad \rho=\phi\rho',\quad\si=\phi\si'$ and
$\Psi_0=\phi^2\Psi_0'$.
 
  Dropping the primes, we have in terms of the affine
parameter $r$:
$$D=\partial/\partial r,$$
and
$$\kappa=\e=0\ . \eqno(3.5)$$
 
  We find that a step-by-step process of integrating the field
equations can be launched in this gauge. The key observation is
that Eqs. (2.16a) and (2.16b), together with (NP~4.2.a) and (NP
4.2.b), {\it i.e.},
$$\Psi_0=-\si (2Dln\phi+\rho-\bar\rho)$$
$$2(\rho+\bar\rho)Dln\phi=(\rho+\bar\rho)^2-2(\rho\bar\rho+\si\bar\si)$$
$$D\rho=\rho^2+\si\bar\si$$
$$D\si =(\rho+\bar\rho)\si+\Psi_0 , \eqno(3.6)$$
form a closed set of equations.
Introducing the real functions $x,y$ and $z$ by
$$x=\rho+\bar\rho \quad\quad y=\rho\bar\rho \quad\quad
z=\si\bar\si,                                      \eqno(3.7)$$
we obtain the autonomous system
$$\eqalign{&Dx=x^2+2(z-y)\cr
&Dy=x(y+z)\cr
&xDz=4z(y+z)\ .} \eqno(3.8)$$
 
   Solution of Eqs. (3.8), by taking the $D$ derivatives and
decoupling, yields the spin coefficients
$$\rho=-{1\oo 2r}\Bigl[1+cos\eta {r^{cos\eta}-iB\oo
r^{cos\eta}+iB} \Bigr] \eqno(3.9)$$
$$\si=-{sin\eta\oo 2r}{r^{cos\eta}+iB\oo r^{cos\eta}-iB}\ .
\eqno(3.10)$$
Here one of the constants has been eliminated by the appropriate
choice
of the origin of the affine parameter $r$. Also we used the remaining
spatial rotations (2.12) to eliminate the $r$-independent part
of the phase factor of $\si$.
The integration functions $\eta$ and $B$ may depend on
the coordinates $(u,x^2,x^3)$. While $B$ can take
any real value, $\eta$ ranges in the
interval $\left[0,360^\circ\right)$.
 
  For the value $B=0$, the null congruence with tangent $l$ is
twist-free, and for $\eta=0$ or $\eta=180^\circ$, it
is shear-free. When both $B=0$ and $\eta=180^\circ$ holds, there
is no expansion.
 
   From (3.6), (3.9) and (3.10) we get the curvature quantity
$$\Psi_0=-{sin2\eta\oo 4r^2}\eqno(3.11)$$
and for $r\ge0$, the scaling function can be written
$$\phi= A^2\Bigl({r^{cos\eta}\oo r^{2cos\eta}+B^2}\Bigr)^{1\oo 2}
\eqno (3.12)$$
where $A$ is another integration function of $u,x^2$ and $x^3$.
 
  We have yet the freedom of performing
the null rotations (2.13). We use these to set
$$\pi=\al+\bar\be \ , \eqno(3.13)$$
removing thereby the term with $D$ from the commutator
$[D,\de]$.
There remain further null rotations (2.13) compatible with (3.13)
where $E$ satisfies
$$DE=\bar\rho E+\si\bar E \ .\eqno(3.14)$$
 
  Let us denote the components of the complex vector $m$ by
$$m=\Omega {\d\oo\d r}+m^i{\d\oo\d x^i}\qquad x^i=u,x^2,x^3 \ .
\eqno(3.15)$$
The second commutator in (NP 4.4), when applied to the
coordinates, yields the equations:
$$D\Omega=\bar\rho \Omega+\si\bar \Omega \eqno(3.16)$$
$$Dm^i=\bar\rho m^i+\si\bar m^i . \eqno(3.17)$$ Noticing that
equations (3.14) and (3.16) have identical forms, we
use the remaining null rotations to arrange
$$\Omega=0 \ .\eqno(3.18)$$
We next integrate Eq. (3.17) to obtain the rest of the components
of $m$:
 $$m^j={1\oo\sqrt r}{1\oo r^{cos\eta}-iB}
\Bigl[Q_1^j r^{cos\eta-\sin\eta\oo2}+iQ_2^j r^{cos\eta+\sin\eta\oo2}
\Bigr],\quad j=1,2,3 \eqno(3.19) $$
where $Q_1^j$ and $Q_2^j $ are again real functions of
$u,x^2,x^3$.

\bigskip
{\bf 4. THE MAIN THEOREM}
 
   We now rewrite the remaining Kerr-Schild equations (2.16.d)
and (2.16e) in the affine gauge. Eq. (2.16.d) takes the
form$^{18}$,
using (NP~4.2.c), (NP 4.2.k), (2.14), (3.2) and (3.4):
$$\de\Bigl({\Psi_0\oo2\si}\Bigr)+{\Psi_0\oo\si}\de
ln\phi-2\si\bar\de ln\phi-
\Psi_1={\Psi_0\oo2\si}(\ta-\bar\al-\be)-2\si(\bar\ta-\al-\bar\be)
+\bar\ta\si-\ta\rho \ .\eqno(4.1)$$
Note that the $\si \to0$ limit is well-behaved as
follows from the relation ${\Psi_0\oo2\si}=-{1\oo2r}-\rho$.
From (2.16e), with (NP 4.2.f), (NP 4.2.q),
(2.14),(3.2) and (3.4):
$$\eqalignno{\de(\bar\ta-\al-\bar\be)+&\bar\de(\ta-\bar\al-\be)-
(\de\bar\de+\bar\de\de)ln\phi+(\rho+\bar\rho)\De ln\phi\cr
&+(2\bar\ta-3\al-5\bar\be)\de ln\phi+(2\ta-3\bar\al-5\be)\bar\de
ln\phi-4(\de ln\phi)(\bar\de ln\phi)=\cr
&2 Re\Psi_2
-(\ga+\bar\ga)(\rho+\bar\rho)+{1\oo2}(\mu+\bar\mu)({\Psi_0\oo\si}
+\rho-\bar\rho)-(\mu-\bar\mu)(\rho-\bar\rho)\cr
&+\ta(\bar\ta-\al-3\bar\be)+\bar\ta(\ta-\bar\al-3\be)\cr
&+4(\al+\bar\be)(\bar\al+\be)-(\al+\bar\be)(\bar\al-\be)
-(\al-\bar\be)(\bar\al+\be)\ .&(4.2)}             $$
 
   Using (3.13) in (NP 4.2.c), (NP 4.2.d), (NP 4.2.e), and
the first Bianchi equation (NP 4.5), we have
$$\eqalignno{
D\ta&=\rho(\ta+\bar\pi)+\si(\bar\ta+\pi)+\Psi_1  &(4.3a)\cr
D\pi&=2\rho\pi+2\bar\si\bar\pi+\bar\Psi_1        &(4.3b)\cr
D\al&=\rho(\pi+\al)+\bar\si(\bar\pi-\bar\al)     &(4.3c)\cr
D\be&=\bar\rho\be+\si(2\pi-\bar\be)+\Psi_1       &(4.3d)\cr
D\Psi_1&=4\rho\Psi_1+(\bar\de+\pi-4\al)\Psi_0.   &(4.3e)}$$
This is a set of coupled linear, inhomogeneous equations for the
functions $\ta,\pi,\al,\be$ and $\Psi_1$ and their complex
conjugates, with the constraints (3.13) and (4.1), algebraic in $r$.
Furthermore, we need Eq. (NP 4.2k),
$$\de\rho-\bar\de\si=\rho\bar\pi+\si(\pi-4\al)+(\rho-\bar\rho)\ta-\Psi_1\eqno(4.4)$$
for the proof of our
 
  \item{ }{\bf Theorem:} For a generalized vacuum-vacuum
Kerr-Schild pencil, either of the following conditions hold:
  \itemitem {(i)} The parameter $\eta$ assumes one of the
special values given by
$$sin\eta=0,\ ^+_-1,\ ^+_-2^{-{1\oo 2}}\quad \ .\eqno(4.5)$$
  \itemitem {(ii)} The $\de$ derivatives are restricted by
$$\de\rho=\de\bar\rho=\de\si=\de\bar\si=\de\Psi_0=\de\phi=0\ .
\eqno(4.6)$$
 
  We prove the theorem by taking the $D$ derivative of both sides
of (4.1) and using Eqs. (4.3),
(3.9), (3.10), (3.11) and the second commutator (NP 4.4) for
eliminating
the $D$ derivatives. We remove $\pi-4\al$ from (4.3e) by Eq.
(4.4). All the unknown spin coefficients cancel and we
arrive at an equation of the form $a\de\phi+b\bar\de\phi=0$.
Eliminating next $\bar\de\phi$ with the help of the complex
conjugate equation, we get
$${cos\eta(r^{2cos\eta}+B^2)+r^{2cos\eta}-B^2\oo
r^{10}(r^{2cos\eta}+B^2)}P\, \de\phi=0      \eqno (4.7)$$
where $P$ is the polynomial in $sin\eta$ and
$cos\eta$
$$P=(2sin^2\eta-1)sin^3\eta\,cos\eta       \eqno(4.8)$$
with roots given in (i). For other values of $\eta$, equation
(4.7) together with
(3.12) yields $\de A=\de B=\de\eta=0$. Using this information,
part (ii) of the theorem is proved.
 
   Condition (4.7) is trivially satisfied in the absence of shear,
$sin\eta=0$. Thus the Kerr solution
will {\sl not} emerge as a limiting case of the shearing
solutions.
 
  Setting aside case (i), the parameter $\eta$ may take real values.
At the next step of the integration process, we are confronted
with the set of coupled homogeneous linear equations
(4.3) for the field quantities $\al,\be,\pi,\ta,\Psi_1$ and their
complex conjugates.
Two linear algebraic relations among these quantities follow from
the adopted gauge, $\pi-\al-\bar\be=0$, and from the Kerr-Schild
condition (4.1):
$$\Psi_1={\Psi_0\oo2\si}(\bar\pi-\ta)+\si(\bar\ta-2\pi)+\rho\ta.\eqno(4.9)$$
Equation (4.4) yields a further algebraic constraint:
$$4\si\al=\si(3\pi-\bar\ta)+\Bigl(\rho-{\Psi_0\oo2\si}\Bigr)\bar\pi
+\Bigl({\Psi_0\oo2\si}-\bar\rho\Bigr)\ta\ .\eqno(4.10)$$
Eqs. (4.3c), (4.3d) and (4.3e) are a consequence of the algebraic
relations and Eqs. (4.3a) and (4.3b).
 
   One can make use of the algebraic constraints to obtain
equations for various closed subsets of the unknown functions. For
example,
using the Kerr-Schild constraint (4.9) for eliminating $\Psi_1$ in
(4.3a) and (4.3b), we
get a set of four coupled equations for $\pi,\ta$:
$$\eqalign{&
D\ta=(2\rho-{\Psi_0\oo2\si})\ta+\si(2\bar\ta-\pi)
                +(\rho+{\Psi_0\oo2\si})\bar\pi\cr
& D\pi=(2\rho+{\Psi_0\oo2\bar\si})\pi+\bar\si\ta
        + (\bar\rho-{\Psi_0\oo2\bar\si})\bar\ta} \eqno(4.11)  $$
and their complex conjugates.
Eqs. (4.3c), (4.3d) and (4.3e) are a consequence of the algebraic
relations and Eqs. (4.11). Therefore, the knowledge of the solution
of Eqs. (4.11) suffices for determining the solution of the
complete system (4.1), (4,3) and (4.4).
 
 Alternatively, Eqs. (4.3b), (4.3c) and (4.3e) form a closed set
for $\al,\pi,\Psi_1$ and their complex conjugates.
 
  Systems of homogeneous linear differential equations always
have nontrivial solutions$^{19}$ in the domain of continuity of
the coefficients. The general solution of $n$ coupled equations
is given by $n$ linearly independent solution vectors.
We will not attempt to tackle this
problem here, but in the next section we shall seek for
space-times characterised by the trivial solution.
 
\bigskip
{\bf 5. A PARTICULAR SOLUTION}
 
  The purpose of this section is to demonstrate that
the class of vacuum space-times
(2.1) with nonvanishing shear $\si$ is not empty. It will suffice
for us to take the trivial solution of Eqs. (4.3):
  $$\ta=\pi=\al=\be=\Psi_1=0\ .\eqno(5.1)$$
 
  Equations (4.9) and (4.10) are identically satisfied. Eqs. (NP
4.2.g), (NP 4.2.h)
and the second relation of (NP 4.5) form a closed system of
linear homogeneous equations for
the field quantities $\la,\mu$ and $\Psi_2$:
$$\eqalignno{&D\la=\rho\la+\bar\si\mu&(5.2a)\cr
&D\mu=\bar\rho\mu+\si\la+\Psi_2&(5.2b)\cr
&D\Psi_2=3\rho\Psi_2-\la\Psi_0\ .&(5.2c)}$$
Equations (NP 4.2.f) and (NP 4.2.l) involve the spin
coefficient $\ga$,
$$\eqalign{&D\ga=\Psi_2\cr
&\Psi_2=\mu\rho-\la\si+\ga(\rho-\bar\rho).}\eqno(5.3)$$
When $B\neq0$, we may use the second equation (5.3) to obtain
$\ga$. The first equation is compatible with this.
The curl-free fields with $B=0$ will be considered elsewhere.
 
In the case when one of the quantities
$\la,\mu,\Psi_2,\ga$
vanishes, Eqs. (5.2) and (5.3) imply that the remainder
must also vanish. (The general solution of the system will
be given in Paper II.)
From (NP 4.2.o) $\nu=0$, from (NP
4.2.i) $\Psi_3=0$ and from (NP 4.2.j) we get $\Psi_4=0$.
Let us choose again this simplest case:
 $$\la=\mu=\ga=\nu=\Psi_2=\Psi_3=\Psi_4=0.\eqno(5.4)$$
 
  From the commutators (NP 4.4) we obtain that $\De$ commutes both
with $\de$ and $D$. Thus we can adopt such coordinates that
$\De={\d\oo\d u}$, $Q_1^1=Q_1^1(x^2,x^3)$ and
$Q_2^1=Q_2^1(x^2,x^3)$.
The functions $A,B$ and $\eta$  are, in fact, constants in this
case because we
already have $\de A=\de B=\de\eta=0$, the fifth relation of (NP 4.5)
implies $\De\Psi_0=0$ or $\De\eta=0$ and from (4.2) we have $\De
ln\phi=0$ or $\De A=\De B=0$. Furthermore
$Q_1=Q_1^j {\d\oo\d x^j}$ and $Q_2=Q_2^j {\d\oo\d x^j}$
are noncommutative but they commute with $\De$ and $D$.
This enables us to choose the coordinates such that
$$Q_1^j=(cos\eta\,Bx^2,0,1)$$
$$Q_2^j=(0,1,0)\ .\eqno(5.5)$$
 
By use of the completeness relation
$$g^{ab}=l^a n^b+n^a l^b-m^a {\bar m}^b-{\bar m}^a m^b\ ,
\eqno(5.6)$$
we obtain the inverse metric
$$g^{ab}=\left(\matrix
       {0&      1&       0&          0&\cr
        1&-hQ_1^2&       0&       -hQ_1\cr
        0&      0&-hr^{2\sin\eta}&    0\cr
        0&  -hQ_1&       0&          -h} \right)
\eqno(5.7)$$
where
$$h(r)=2{r^{cos\eta-sin\eta-1}\oo r^{2cos\eta}+B^2}
\qquad Q_1=cos\eta Bx^2\ .$$
 
Hence we get the metric
$$d\tilde s^2=ds^2-
V_o{r^{cos\eta}\oo
r^{2\cos\eta}+B^2}(du-cos\eta\,Bx^2dx^3)^2\eqno(5.8)$$
where $V_0=\la A^4$ is a constant and $ds^2$ is the parent
metric of the Kerr-Schild pencil:
$$ds^2=2dr\,du-2cBx^2dr\,dx^3
-{r^{2cos\eta}+B^2\oo2 r^{cos\eta}}\Bigl[
r^{1-sin\eta}(dx^2)^2+ r^{1+sin\eta}(dx^3)^2\Bigr]   \eqno(5.9)$$
 
By performing the Eddington-type coordinate transformation
$$du=dt+{r^{2cos\eta}+B^2\oo V_or^{cos\eta}}dr\ ,\eqno(5.10)$$
the metric assumes the form
$$\eqalign{d\tilde s^2&=
-V_o{r^{cos\eta}\oo r^{2\cos\eta}+B^2}(dt-cos\eta\,Bx^2dx^3)^2 \cr
&-{r^{2cos\eta}+B^2\oo2 r^{cos\eta}}\Bigl[{-2\oo V_o}  dr^2
+ r^{1-sin\eta}(dx^2)^2+ r^{1+sin\eta}(dx^3)^2\Bigr]. }
\eqno(5.11)$$
 
 This is one of the K\'ota-Perj\'es$^{20}$ vacuum solutions, and thus
(5.11) admits a Killing vector $\partial/\partial
t$ whose {\it eigenrays} are geodesic.
 
\medskip
{\bf 6. CONCLUDING REMARKS}
 
   The picture we glean in this work is as follows. The vacuum
space-times in which
vacuum Kerr-Schild metrics can be generated are characterized
either (i) by special values of the parameter $\eta$, or by
condition
(ii) of our theorem. The metrics in class (ii) are a
dynamical system governed by two sets of homogeneous linear
equations (4.3), with $\bar\de\Psi_0=0$, and (5.2). The
spectrum of the system is constrained by
the nonradial Newman-Penrose and Kerr-Schild equations. The ground
state is a K\'ota-Perj\'es space-time. At the moment of writing it
appears to us that the excited states can all be
generated in terms of elementary functions. This issue will be
discussed in Paper II.
\medskip
{\bf ACKNOWLEDGMENT}
\smallskip
   We thank Professor Roy P. Kerr for providing inspiration for
our work.
{\ \ }
\medskip
{\bf REFERENCES}
\frenchspacing
\smallskip
\item{[1]} Kerr, R. P.  and Schild, A., {\it Atti Del
Convegno Sulla Relativit  Generale: Problemi Dell' Energia e Onde
Gravitazionali (Anniversary Volume, Forth Centenary of Galileo's
Birth)}, G. Barb'ra, Ed. (Firenze, 1965), p. 173,
and Trautman, A., in {\it Recent Developments in General
Relativity}, Pergamon Press, p. 459 (1962)
\item{[2]} Boyer, R. H. and Lindquist. R. W.,
J. Math. Phys. {\bf8}, 256 (1967)
\item{[3]} Grses, M. and Grsey, F.
J. Math. Phys. {\bf16}, 2385 (1975)
\item{[4]} Debever, R. , Bull. Acad. Roy. Belgique Cl. Sci.
{\bf60}, 998 (1974)
\item{[5]} McIntosh, C.B.G.: Kerr-Schild Spacetimes Revisited,
in {\it Conference on Mathematical Relativity}, Ed. R. Bartnik,
Proc. Centre for Mathematical Analysis,
Australian National University, Canberra, 1988.
\item{[6]} McIntosh, C.B.G. and Hickman, M.S., Gen. Rel. Grav.
{\bf 20}, 793 (1988)
\item{[7]} Chandrasekhar, S.: The Mathematical Theory of Black
Holes, Oxford Univ. Press, Oxford, 1983, Chap.57.
\item{[8]} Debney, G. C., Kerr, R. P.  and Schild,
A. , J. Math. Phys. {\bf10}, 1842 (1969)
\item{[9]} D. Kramer {\it et al.}: {\it Exact Solutions of
Einstein Field Equations}, Cambridge Univ. Press (1980)
\item{[10]} Thompson, A.H., Tensor {\bf17}, 92 (1966)
\item{[11]} Dozmorov, I.M., Izv. VUZ. Fiz.{\bf11}, 68 (1971)
\item{[12]} What Thompson really
proves is that the image of the Kerr-Schild map from vacuum to
vacuum is algebraically
special {\sl provided} the parent space-time is algebraically
special, and the Kerr-Schild null vector $l$ is a multiple
principal vector
\item{[13]} Xanthopoulos, B. C., J. Math. Phys. {\bf19}, 1607 (1978)
\item{[14]} H. Urbantke, Acta Physica Austriaca {\bf 41},1 (1975)
\item{[15]} E. Nahmad-Achar, J.Math.Phys. {\bf29},1878 (1988)
\item{[16]} Wald, R.M.: {\it General Relativity}, The University
of Chicago Press (1984)
\item{[17]} Newman, E. and Penrose, R., J. Math. Phys. {\bf3}, 566
(1962)
\item{[18]} Equations (4.1) and (4.2) are given here in a form
which is valid with no regard to the gauge condition (3.13)
\item{[19]} Kamke, E.: {\it Differentialgleichungen -
L"sungsmethoden und L"sungen I.}, Sec. 8.2. Teubner, Stuttgart,
1977
\item{[20]} K\'ota, J. and Perj\'es, Z., J. Math. Phys. {\bf13}, 1695
(1972)
\end